\documentclass[prd,twocolumn,a4paper,superscriptaddress]{revtex4}
\usepackage{graphicx,amssymb}
\begin{document}
%My commands
\newcommand{\be}{\begin{equation}}
\newcommand{\ee}{\end{equation}}
\newcommand{\bq}{\begin{eqnarray}}
\newcommand{\eq}{\end{eqnarray}}
\newcommand{\bsq}{\begin{subequations}}
\newcommand{\esq}{\end{subequations}}
\newcommand{\bc}{\begin{center}}
\newcommand{\ec}{\end{center}}
\newcommand {\R}{{\mathcal R}}
\newcommand{\al}{\alpha}
\newcommand\lsim{\mathrel{\rlap{\lower4pt\hbox{\hskip1pt$\sim$}}
    \raise1pt\hbox{$<$}}}
\newcommand\gsim{\mathrel{\rlap{\lower4pt\hbox{\hskip1pt$\sim$}}
    \raise1pt\hbox{$>$}}}

\title{Is $w \neq -1$ evidence for a dynamical dark energy equation of state ?}
\author{P. P. Avelino}
%\email[Electronic address: ]{ppavelin@fc.up.pt}
\affiliation{Centro de F\'{\i}sica do Porto, Rua do Campo Alegre 687, 4169-007 Porto, 
Portugal}
\affiliation{Departamento de F\'{\i}sica da Faculdade de Ci\^encias
da Universidade do Porto, Rua do Campo Alegre 687, 4169-007 Porto, 
Portugal}
\author{A. M. M. Trindade}
%\email[Electronic address: ]{arlindo.trindade@gmail.com}
\affiliation{Centro de Asrof\'{\i}sica da Universidade do Porto, Rua das Estrelas 687, 4150-762 Porto, 
Portugal}
\author{P. T. P. Viana}
%\email[Electronic address: ]{viana@astro.up.pt}
\affiliation{Centro de Asrof\'{\i}sica da Universidade do Porto, Rua das Estrelas 687, 4150-762 Porto, 
Portugal}
\affiliation{Departamento de Matem\'atica Aplicada da Faculdade de Ci\^encias
da Universidade do Porto, Rua do Campo Alegre 687, 4169-007 Porto, 
Portugal}

\date{29 June 2009}
\begin{abstract}

Current constraints on the dark energy equation of state parameter, $w$, are expected to be improved by more than one order of magnitude in the next decade. If $|w-1| \gsim 0.01$ around the present time, but the dark energy dynamics is sufficiently slow, it is possible that future constraints will rule out a cosmological constant while being consistent with a time-independent equation of state parameter. In this paper, we show that although models with such behavior can be constructed, they do require significant fine-tuning. Therefore, if the observed acceleration of the Universe is induced by a dark energy component, then finding $w \neq -1$ would, on its own, constitute very strong evidence for a dynamical dark energy equation of state.

\end{abstract}
\maketitle

\section{Introduction}

The observational evidence for an acceleration of the expansion of the Universe in the recent past is now overwhelming (see for example \cite{Frieman:2008sn} for a review), but the precise cause of this phenomenon is still unknown. The most popular scenario assumes that such acceleration is the result of the existence, in the Universe, of a nearly homogeneous dark energy component violating the strong energy condition, described by a minimally coupled scalar field. If the scalar field is static then it will  give rise to a non-zero vacuum energy density, also known as a cosmological constant. However, given the enormous discrepancy between the observationally inferred vacuum energy density and theoretical expectations, a dynamical scalar field is expected to be a more plausible explanation.

The dark energy density does not need to be homogeneous. In fact, a number of inhomogeneous dark energy models have been proposed in the literature. For example, domain wall networks have been proposed as an alternative explanation to the present acceleration of the Universe \cite{Bucher:1998mh} although recent results seem to exclude this possibility \cite{PinaAvelino:2006ia,Avelino:2008ve}. Another example is provided by unified dark energy models where dark matter and dark energy are strongly coupled to each other and behave as a single fluid (see for example \cite{Avelino:2008cu}). Other possibilities include modifications to General Relativity \cite{Carroll:2004de,Copeland:2006wr}, like those associated with extra-dimensions or modifications to the coupling to spatial curvature, an example being f(R) theories.

Current observations already provide some interesting limits on the equation of state of the dark energy, usually parameterized by $w$, but its dynamics is still poorly constrained \cite{Komatsu:2008hk,Biswas:2009de}. This situation is expected to change in the next decade \cite{Albrecht:2006um,Tang:2008ct,Albrecht:2009ct}. However, it is possible that the constraints may then still be found to be compatible with a time-independent $w$. In this case, it would be important to know to what extent it would be worthwhile to try to tighten further the constraints on $w$ in order to identify possible variations with time. It is this question that we address in the following sections. 

Throughout this paper we use units in which $c=\hbar=8 \pi G=1$.

\section{Generic scalar field models}

In this section we shall consider a broad class of dark energy models described by a single real scalar field 
with action
\begin{equation}\label{eq:L}
S=\int d^4x \, \sqrt{-g} \, {\mathcal L} \, ,
\end{equation}
where ${\mathcal L} (\phi,X)$ is the scalar field Lagrangian, $X=\phi_{, \mu} \phi^{, \mu} /2$ and a comma is used to 
represent a partial derivative.

The energy-momentum tensor of the scalar field may be written 
in a perfect fluid form
\begin{equation}\label{eq:fluid}
T^{\mu\nu}= (\rho+ p) u^\mu u^\nu - p g^{\mu\nu} \,,
\end{equation}
by means of the following identifications
\begin{equation}\label{eq:new_identifications}
u_\mu = \frac{\phi_{, \mu}}{\sqrt{2X}} \,,  \quad \rho = 2 X {\mathcal L}_{,X} - {\mathcal L} \, ,\quad p =  {\mathcal L}(X,\phi)\, .
\end{equation}
In Eq.~(\ref {eq:fluid}), $u^\mu$ is the 4-velocity field describing the motion of the fluid (for timelike $\phi_{, \mu}$), while $\rho$ and $p$ are its proper energy density and pressure, respectively. The dark energy equation of state parameter $w$ is 
\begin{equation} 
\label{eq:w}
w \equiv \frac{p}{\rho} = \frac{\mathcal L}{2X {\mathcal L}_{,X}  - {\mathcal L}}, 
\end{equation} 
and the sound speed squared is given by
\begin{equation}
\label{eq:cs2}
c_s^2 \equiv \frac{p_{,X}}{\rho_{,X}}=\frac{{\mathcal L}_{,X}}{{\mathcal L}_{,X}+2X{\mathcal L}_{,XX}}\, ,
\end{equation}
as long as ${\mathcal L}_{,X}\ne 0$.

We can re-write Eq.~(\ref {eq:w}) as
\begin{equation} 
\label{eq:wapp}
w^{-1}  = -1 + \frac{2X {\mathcal L}_{,X}}{\mathcal L}\,.
\end{equation} 
Further assuming that $w$ is always constant, independently of the value of $\phi$ and $X$, then leads to
\begin{equation} 
{\mathcal L} = f(\phi) X^{(1+w^{-1})/2}\,,
\end{equation} 
where $f(\phi)$ is an arbitrary function of $\phi$. Models with ${\mathcal L} = f(\phi) X^n$ and constant $n$ yield $c_s^2 = 1/(2n-1)$. In particular, if $n = 1$ then $c_s^2=1$, corresponding to a massless scalar field, if $n = 2$ then $c_s^2=1/3$, corresponding to background radiation, and in the $n \to \infty$ limit one has $c_s^2 \to 0$, corresponding to pressureless non-relativistic matter. However, if $n=(1+w^{-1})/2 \sim 0$ then Eq.~(\ref {eq:cs2}) implies that $c_s^2 \sim w \sim -1$. These models have a negative sound speed squared, which would necessarily lead to the development of very large inhomogeneities in the spatial distribution of the dark energy density, strongly disfavored by the observational data. Therefore, the only realistic way in which $w$ can be a constant, irrespectively of the value of $\phi$ and $X$, is to be exactly equal to $-1$.

\section{Homogeneous dark energy}

It is possible to construct dark energy models where $w$ is as close as desired to a constant within some time interval, but different from $-1$, and without instabilities being generated. However, we will show that such models are in general very contrived. 

Consider, for example, quintessence dark energy models described by a real scalar field 
with Lagrangian 
\begin{equation} 
{\cal L}= X - V(\phi)\,. 
\end{equation} 
Generically, the equation describing the dynamics of a scalar field may be obtained by varying the action with respect to $\phi$
\begin{equation} 
\frac{1}{{\sqrt {-g}}} \left({\sqrt {-g}} {\mathcal L} _{,X}  \phi^{, \mu} \right)_{,\mu}={\mathcal L}_{,\phi}\,. 
\end{equation} 
Assuming a flat Friedmann-Robertson-Walker metric with line element
\begin{equation} 
ds^2=dt^2 - a^2(t)\left(dx^2+dy^2 +dz^2\right) \,,
\end{equation} 
and a (nearly) spatially homogenous dark energy component, the scalar field equation of motion is approximately given by
\begin{eqnarray}
\frac{\partial}{\partial t}\left({\mathcal L}_{,X}  \frac{\partial \phi}{\partial t}\right) + 3H{\cal L}_{,X}  \frac{\partial \phi}{\partial t} = {\cal L}_{,\phi}\label{eq:sfdyn}\,.
\end{eqnarray}
which, by introduction of the proposed Lagrangian, reduces to
 \begin{equation} 
\label{eq:motphi} 
{\ddot \phi}+3H {\dot \phi} = -V_{,\phi}\,,
\end{equation} 
where a dot represents a derivative with respect to the physical time, $t$. The dark energy equation of state parameter $w$ is given by
 \begin{equation} 
w \equiv \frac{p}{\rho}=\frac{ {\dot \phi}^2/2-V(\phi)}{{\dot \phi}^2/2+V(\phi)}\label{eq:wxv}\,.
\end{equation} 
and the sound speed squared is $c_s^2=1$. The fact that the sound speed is equal to the speed of light prevents the generation of large spatial fluctuations in the dark energy density. 
If we require $w$ to be a constant then Eq.~(\ref {eq:wxv}) implies
 \begin{equation} 
  \label{eq:ddotphi} 
 {\ddot \phi}= V_{,\phi} \frac{1+w}{1-w}\,,
\end{equation} 
and
 \begin{equation} 
 \label{eq:dotphi} 
{\dot \phi}= \pm \left(2 V \frac{1+w}{1-w}\right)^{1/2}\,.
\end{equation} 

Note that if $w \sim -1$ then $| \ddot \phi | \ll |V_{,\phi}|$.
In the following we shall drop the $\pm$ sign. It will be suficient to realize for each solution with $\dot \phi > 0$ and $V_{,\phi} < 0$ 
there will be another one with $\dot \phi < 0$ and $V_{,\phi} >0$. From now on we shall only consider the solutions with $\dot \phi > 0$.
 
Substituting Eqs.~(\ref {eq:ddotphi}) and (\ref {eq:dotphi})  into Eq. (\ref {eq:motphi}) one 
obtains
 \begin{equation} 
 \label{eq:potev}
\frac{(V_{,\phi})^2}{V}=\frac{9}{2}(1-w^2) H^2\,.
\end{equation}
For this equation to be verified the potential of $\phi$ would have to be designed such that $V_{,\phi}^2/V \propto H^2$. This requires very large fine-tuning, in particular during the transition from the matter to the dark energy dominated eras, as we shall see next.

Multiplying equation (\ref {eq:potev}) by ${\dot \phi}^2$ and using equation (\ref {eq:dotphi}) it is simple to show that
\begin{equation} 
\label{eq:pota} 
V=V_0 a^{-3(1+w)}\,,
\end{equation}
where the subscript `0' refers to the present time (we are taking $a_0=1$). 
However, the evolution of $\phi$ with the scale factor $a$ is, in general, very different in the matter and dark energy dominated eras. 
In fact, assuming a flat universe, one has
\begin{equation} 
H^2=H_ 0^2\left(\Omega_{m0}a^{-3}+\Omega_{e0}a^{-3(w+1)}\right)\,,
\end{equation}
so that, using Eq. (\ref {eq:dotphi}), one obtains
\begin{equation}
\label{eq:dphida}
\frac{d\phi}{da}= {\sqrt {3\Omega_{e0}(1+w)}}\left(\Omega_{m0}a^{3w+2}+\Omega_{e0} a^2\right)^{-1/2}\,,
\end{equation}
which has the solution
\begin{equation}
\label{eq:sdphida}
\phi= A + B \ln \left(\frac{a^{3w}}{\left(1+\left(\Omega_{m0}a^{3w}/\Omega_{e0}+1\right)^{1/2}\right)^2}\right)\,,
\end{equation}
where $B= {\sqrt {3(1+w)}}/(3w)$, $A$ is an arbitrary integration constant, $\Omega_{m 0}=\rho_{m 0}/(3H_0^2)$ and  $\Omega_{e 0}=\rho_{0}/(3H_0^2)$ (note that $\rho$ is the dark energy density). In this paper we take $\Omega_{m0}=0.27$ and $\Omega_{e 0}=0.73$ as  favored by the five-year WMAP results \cite{Komatsu:2008hk}.

At very late times ($a \gg 1$) the dark energy will completely dominate the energy density of the universe, and the evolution of $\phi$ with the scale factor will be given by
 \begin{equation}
 \label{eq:c1c2}
\phi=C_e+{\sqrt{3(1+w)}} \ln a\,,
\end{equation}
where $C_e$ is an arbitrary constant.
Using Eq. (\ref {eq:pota}) one obtains the following solution
\begin{equation} 
\label{eq:ve}
V=V_e \exp\left(-{\sqrt{3 (1+w)}}(\phi-\phi_e)\right)\,,
\end{equation}
valid at an arbitrary time $t_e$ well into the dark energy dominated era (to which the subscript `$e$' refers to).

On the other hand, at early times ($a \ll 1$) deep into the matter era one has
\begin{equation} 
\label{eq:c3c4}
\phi= C_m-\frac{2}{3w} {\sqrt \frac{3(1+w)\Omega_{e0}}{\Omega_{m0}}} a^{-3w/2}\,,
\end{equation}
where $C_m$ is an arbitrary integration constant. Using equation (\ref {eq:pota}) one obtains the following solution
\begin{equation} 
\label{eq:vm}
V \propto (\phi-C_m)^{2(w+1)/w}\,,
\end{equation}
valid deep into the matter era.

\begin{figure}[t!]
\includegraphics[width=7.5cm]{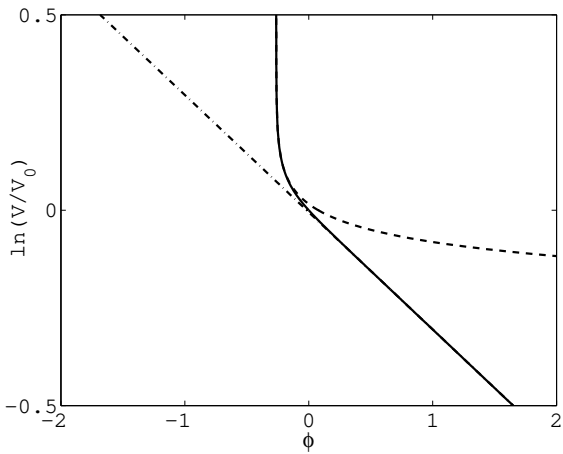}
\caption{\small{The solution for $V(\phi)$ assuming that $w_0=-0.97$ at all times (solid line), as well as the analytical solutions for the scalar field potential, computed using Eqs. (\ref {eq:vm}) or  (\ref {eq:ve}), valid deep into the matter era (dashed line) and dark energy era (dot-dashed line) respectively.}}
\end{figure}

\begin{figure}[t!]
\includegraphics[width=7.5cm]{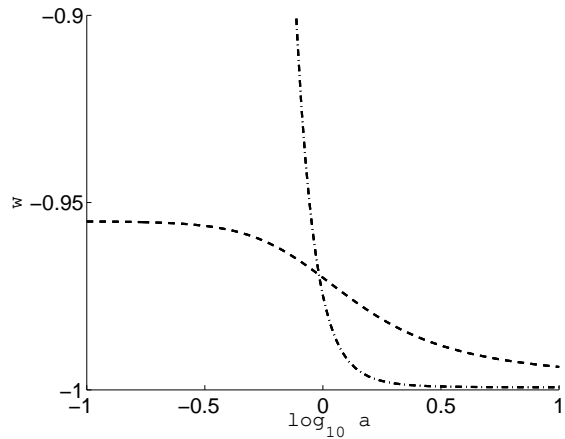}
\caption{\small{The evolution of the equation of state parameter computed with the potentials given in Eqs. (\ref {eq:vm}) or  (\ref {eq:ve}) (dashed and dot-dashed lines, respectively).
}}
\end{figure}

In Fig.~1 we plot  the solution for $V(\phi)$ assuming that $w_0=-0.97$ at all times (solid line), as well as the analytical solutions, computed using Eqs. (\ref {eq:vm}) or  (\ref {eq:ve}), valid deep into the matter and dark energy eras (dashed and dot-dashed lines, respectively). The initial conditions for the $w={\rm constant}$ solution were chosen so that $\phi_0=0$ and the constants $C_m$ and $C_e$ were determined by requiring that the analytical solutions computed using Eqs. (\ref {eq:vm}) or  (\ref {eq:ve}) fitted the constant $w$ results obtained deep into the matter and dark energy dominated eras, respectively. 
It is clear from Fig.~1 that, in order to obtain $w={\rm constant}$, the shape of the potential must be fine-tuned around $\phi=\phi_0$. Otherwise,  the equation of state parameter would change rapidly around the present time. This can be seen in Fig.~2, where we plot the evolution of the equation of state parameter with the potentials given by Eqs. (\ref {eq:vm}) or  (\ref {eq:ve}) (dashed and dot-dashed lines, respectively). These potentials, designed to produce a constant $w$ deep into the matter and 
dark energy dominated eras, respectively, give rise to a rapidly changing $w$ in the transition between them, with 
$|w_0-w(z=1)|/|w_0+1| \gsim 1$ (here $z=1/a-1$ is the redshift).

If the scalar field is in slow-roll then Eq. ~(\ref {eq:sfdyn}) is reduced to 
\begin{equation} 
3 H  {\mathcal L}_{,X} (2 X)^{1/2} =  {\mathcal L}_{,\phi} \,,
\end{equation}
with $X={\dot \phi}^2/2$ and
\begin{eqnarray} 
H^2 &=&  \left(\rho_{m 0} a^{-3} + 2 X {\mathcal L}_{,X} - {\mathcal L}\right)/3\nonumber\\
&=&  \left(\rho_{m 0} a^{-3} + w^{-1}{\mathcal L}\right)/3\,.
\end{eqnarray}
Hence, deep into the dark energy era, one has
\begin{equation} 
\label{eq:c6}
\left(\frac{X}{|\mathcal L|}\right)^{1/2} (\ln |{\mathcal L}|)_{,\phi} = {\sqrt \frac32} |w|^{-1/2} (1+w^{-1})\,.
\end{equation} 
assuming a constant $w$. Thus, given $\phi_e$, $X_e$,  ${\mathcal L} (\phi_e,X_e)$ and $w$ one is able to construct ${\mathcal L} (\phi,X)$ using Eqs.~(\ref {eq:wapp}) and  (\ref {eq:c6}) deep into the dark energy dominated era. Of course, the extension of this solution to the matter dominated era or to the transition between matter and dark energy dominated eras would necessary imply a varying equation of state parameter. This  happens because, in order to describe the 
dark energy, the evolution of ${\mathcal L}$ with the scale factor around the present time must be much slower than that of the matter density.

\section{Inhomogeneous models}

\subsection{Topological defects}

Cosmic defect networks can usually be characterized by two very different characteristic scales. One is the defect thickness, $\delta$, which is determined by the particle physics model. In general, it remains constant in physical coordinates, in which case it is insensitive to the large scale dynamics of the Universe. The other characteristic scale, $L$, is associated with the large scale properties of the network. This scale is affected by the  cosmology and is proportional to the scale factor, $a$, for frozen defects. This is in fact the most interesting solution from a dark energy point of view since it corresponds to the case where the evolution of the average defect density with redshift is slower. In this case, the (average) defect energy density, $\rho$, is given by 
\begin{equation} 
\rho \propto a^{-N}\,,
\end{equation}
where $N$ is defect's spatial dimension ($N=0,1,2$ respectively for  point masses, strings or domain walls). If the defects are minimally coupled to all other fields then energy-momentum conservation implies an (average) equation of state parameter 
equal to $w=-N/3$. However, if the defects have a non-zero root mean square velocity, $v$, then the (average) equation of state parameter is given by
\begin{equation} 
w=-\frac{N}{3}+C(N)v^2\,.
\end{equation}
where $C(N)=(N+1)/3$ so that $w \to 1/3$ for  $v \to 1$. In order to accelerate the expansion of the universe it is necessary that 
$w  < -(1+\Omega_{m0}/\Omega_{e 0})/3$ and consequently only a frustrated domain wall network could in principle do the job. However, a combination of analytical and numerical results have provided very strong evidence that no frustrated domain wall network, with $v \sim 0$ and $L \propto a \ll H^{-1}$, is ever expected to arise from realistic initial conditions, invalidating domain walls as a viable dark energy candidate \cite{PinaAvelino:2006ia, Avelino:2008ve}. Furthermore, current observational constraints on the equation of state parameter of dark energy already strongly disfavor $w=-2/3$ \cite{Komatsu:2008hk}.

\subsection{Unified dark energy models}

Although dark matter and dark energy are usually treated as separate components 
minimally coupled to each other, this does not need to be the case (see for example \cite{Brax:2004qh}). In particular, if 
a strong coupling between dark matter and dark energy exists then they may behave as a single fluid. The best known example is provided by the (generalized) Chaplygin gas \cite{Bilic:2001cg,Bento:2002ps} where dark matter and dark energy are described by a single perfect fluid with equation of state parameter:
\begin{equation}
w=-A/\rho^{1+\alpha}\,,
\end{equation}
where $\alpha > 0$ is a constant, and a sound speed squared $c_s^2 = - \alpha w$.

If $\alpha=0$ the Chaplygin gas model is exactly equivalent to a $\Lambda$CDM model \cite{Avelino:2003cf} but for other choices of $0 < \alpha \le 1$ it gives rise to a background evolution identical to that of a quintessence model with a variable equation of state parameter.
However, there are other possibilities for the equation of state parameter of unified dark energy, and we can in fact design it so that it mimics the background evolution of a quintessence dark energy model with constant $w$. This can be done by carefully designing the equation of state of the unified dark energy component with density $\rho_u$ and pressure $p_u$ so that
\begin{equation}
\label{eq:rhou}
\rho_u=\frac{p_u}{w}+\rho_{m0}\left(\frac{p_u}{w \rho_{0}}\right)^{1/(1+w)}\,,
\end{equation}
where $\rho_m$ and $\rho$ are the matter and quintessence dark energy densities, respectively.
However, it is simple to show that, if $w > -1$, this gives rise to a negative sound speed squared making the model unstable to linear perturbations \cite{Bean:2007ny,Avelino:2008zz}.

Furthermore, it has been shown that non-linear effects can significantly modify the evolution of the Universe compared to the 
linear expectations \cite{Beca:2007rd,Avelino:2008zz}. The negative sound speed associated with  Eq.~(\ref {eq:rhou}) will make (almost) empty regions even emptier with $p_u \to 0$. Consequently, the transition from the decelerating to the accelerating phase may never happen in this case.

On the other hand, in the case of the (generalized) Chaplygin gas ($c_s^2 >0$) it has been shown that non-linear effects may anticipate the transition from the dark-matter to the dark energy  dominated eras leading to a background evolution very similar to the $\Lambda$CDM model, even for $\alpha \neq 0$. Either way, the coupling  between dark matter and dark energy is not expect to alleviate the fine-tuning associated with a constant $w$.

\section{\label{conc}Conclusions}

We have shown that in order for the dark energy equation of state parameter, $w$, to be constant in time and close to, albeit different from, $-1$, a significant amount of fine-tuning would be required in the wide range of models considered. This is essentially the result of the existence of a transition era between matter and dark energy domination in the recent past. Therefore, any future evidence which excludes $w=-1$, even if it is consistent with a time-independent value for $w$, should be interpreted as indicative of a dynamical dark energy equation of state. Clearly, in that situation, a further tightening of the constraints on the time variation of $w$ should be actively sought.

%Mention the possibility that gravity itself may be generating the observed accelerated expansion of the Universe.

%%%%%%%%%%%%%%%%%%%%%%%%%%%%%%%%%%%%%%%%%%%%%%%%%%%%%%%%%%%

\bibliography{DE}

\end{document}